\documentclass[12pt]{article}
\newcommand{\be}{\begin{equation}}
\newcommand{\ee}{\end{equation}}
\newcommand{\bea}{\begin{eqnarray}}
\newcommand{\eea}{\end{eqnarray}}
\newcommand{\nn}{\nonumber}

\newcommand{\ksi}{\xi}

       \newcommand{\bj}{{\mathbf j}}

       \newcommand{\bB}{{\mathbf B}}

       \newcommand{\bE}{{\mathbf E}}

\newcommand{\bv}{{\mathbf v}}
\newcommand{\ed}{\end{document}}
\usepackage{color}

\usepackage{epsfig}
\usepackage{epstopdf}

\usepackage[a4paper,
            bindingoffset=0.2in,
            left=1in,
            right=1in,
            top=1in,
            bottom=1in,
            footskip=.25in]{geometry}

\usepackage{hyperref}
\hypersetup{
    colorlinks=true,
    linkcolor=red,
    filecolor=magenta,      
    urlcolor=cyan,
    citecolor = cyan,
    }

\begin{document}

\begin{center}
\noindent
{\large \bf New axisymmetric equilibria with flow from

\vspace{2mm}
an expansion about the generalized Solov'ev solution}\vspace{4mm}

A. I. Kuiroukidis$^1$, D. A. Kaltsas$^{1}$ and G. N. Throumoulopoulos$^1$ \vspace{4mm}

$^1$Department of Physics, University of Ioannina, GR 451 10 Ioannina, Greece  \vspace{4mm}

\vspace{3mm}
Emails: a.kuirouk@uoi.gr,\ kaltsas.d.a@gmail.com, \  gthroum@uoi.gr
\end{center}


\begin{abstract}

We construct analytic solutions to the generalized Grad-Shafranov equation, which incorporates both toroidal and poloidal flows. This is achieved by adopting a general linearizing ansatz for the free-function terms of the equation and expanding the generalized Solov'ev solution [Ch. Simintzis, G. N. Throumoulopoulos,
G. Pantis and H. Tasso, Phys. Plasmas {\bf 8}, 2641 (2001)]. On the basis of these solutions, we examine how the genaralized Solov'ev configuration is modified as the values of the  free parameters associated with the additional pressure, poloidal-current
and electric-field terms are  changed. Thus, a variety of equilibria
of tokamak, spherical tokamak and spheromak pertinence are constructed,  including D-shaped configurations with positive and negative triangularity and diverted configurations with either a couple of X-points  or a single X-point.

\end{abstract}



\section{Introduction}\

The magnetohydrodynamics axisymmetric equilibrium states are governed by the well-known Grad-Shafranov (GS) equation, a quasi-linear elliptic partial differential equation  for the poloidal magnetic flux-function, containing a couple of arbitrary functions of that flux (surface functions), namely the pressure and the poloidal-current functions. This equation is generally solved numerically under appropriate boundary conditions, and to this end, several codes have been developed, e.g. the  HELENA code \cite{HELENA1}. Also, to easier gain physical intuition and for benchmarking equilibrium codes, analytic solutions have been constructed to linearized forms of the GS equation. In connection with the present study we first mention the widely employed
in plasma confinement studies Solov'ev solution \cite{sol}. This solution,   corresponding to constant surface-function terms in the GS equation,  describes an up-down symmetric equilibrium with D-shaped magnetic surfaces surrounded by a spontaneously formed separatrix with a couple of X-points (cf. Fig. \ref{Fig 0}). Also, known is the Maschke-Hernegger solution \cite{he,ma}, expressed in terms of the Whittaker functions, corresponding to linear choices of the surface-function terms  and describing more realistic equilibria  with current-density profiles vanishing on the plasma boundary. Several additional analytic solutions to the GS equation are available, e.g.  extending the Solov'ev solution    by polynomial contributions to the homogeneous counterpart of the GS equation in order to construct diverted tokamak equilibria \cite{srila}; for constant pressure  and linear current-density choices of the free terms in the GS equation \cite{maz};
for linear choices of both the  pressure and  current-density  terms by imposing  D-shaped boundaries or  boundaries with a lower X-point 
\cite{gufr,cefr,ceon}; by differentiation of known separated solutions with respect to the separation parameter in order to get simpler solutions \cite{pfre};
and in elliptic prolate geometry \cite{cri}.  Also, for the case of the generic linearized form of the GS equation an analytic solution involving infinite series was derived in \cite{ata} and was employed to construct equilibrium configurations pertinent to the tokamak ASDEX-Upgrade.

Extended equilibrium investigations have been conducted to include  plasma flows and associated electric fields, which play  an important role in the creation of   advanced confinement regimes in tokamaks, e.g. the transition of the low  to high confinement mode. Specifically,  to include axisymmetric toroidal flows, inherently incompressible because of the symmetry,  elliptic PDEs were derived in \cite{mape} for either isentropic or isothermal magnetic surfaces and analytic solutions were constructed therein. For compressible flows of arbitrary direction the problem reduces to a set of a PDE coupled with a Bernoulli equation involving the pressure \cite{sol67,mosol,ham}. Depending on the value of the poloidal velocity,  this PDE can be either elliptic or hyperbolic, that is, there are three critical transition poloidal velocities, thus forming two elliptic  and two hyperbolic regions. In the frame of two-fluid model, the electron and ion surfaces depart from the magnetic surfaces and the problem reduces to a set of three PDEs involving respective flux-function labels \cite{ste,clth}. The problem was also investigated within the framework of simplified  two-fluid models, such as Hall-MHD  that neglects the electron inertia \cite{il2001,itra} and extended MHD \cite{kath2017,kath2018,kath2019}.  The impact of pressure anisotropy was examined in \cite{cl,il1996,zwer,evth,kuev}. In addition, to include fast particle populations the problem was addressed  in the frame of hybrid fluid-kinetic modes  \cite{grad,iabo,kamo2021,kamo2023,kaku2024}.  

For incompressible MHD flows having a poloidal component  the generalized GS equation becomes elliptic (Eq. (\ref{eq1}))  \cite{tath,sim} and decouples from the Bernoulli equation,  which can be employed as a formula for the pressure (Eq. \ref{pr}). Incompressibility, implying that the density becomes uniform on the magnetic surfaces, is an acceptable approximation for fusion laboratory plasmas for two reasons: First, the poloidal velocities of those plasmas lie well within the first elliptic region. Second, since for typical flows in laboratory fusion plasmas, i.e. for Alfv\'en Mach numbers on the order of 0.01, density variations on the magnetic surfaces are very low.    

Equation (\ref{eq1}), which contains five free surface quantities,  has an additional electric-field term  through the electrostatic potential $\Phi$, associated  with the component of plasma velocity non-parallel to the magnetic field. For parallel velocity and arbitrary assignment of the free surface-function terms, an extension of the HELENA code was developed in \cite{HELENA2}, while the generalized Solov'ev solution to (\ref{eq1}) (given by  (\ref{eq4})), which constitutes  a basic ingredient of the present study,  was obtained in \cite{sim}. Owing to the electric field, the respective equilibrium  in addition to a separatrix, similar to that of the usual static Solov'ev equilibrium, possess  a third X-point  outside the separatrix (cf. Fig. 11 of \cite{sim}). Other analytic solutions to the generalized GS equation of tokamak or space plasma pertinence were obtained by adopting a Solov'ev-like linearizing ansatz and imposing a diverted boundary \cite{kath2014}; alternative linearizing ansatzes \cite{shi2011,kuth2016,kaku2019}; non-linear forms of the generalized GS equation \cite{sal2018}; employing Lie-point symmetries for linear and nonlinear choices of the free-function terms \cite{kuth2016a, kuth2014,poth2023}; and by the method of similarity reduction \cite{kath2016,kuka2024}.

Aim of the present work is to construct analytically solutions to the generalized GS equation in its generic linearized form by employing an alternative method. This  consists in pursuing solutions as expanded generalized Solov'ev ones, i.e. as superposition of the generalized Solov'ev solution and a function to be determined. In addition to the novelty of this method we are particularly interested in examining how the Solov'ev configuration is modified when the influence of the expanding function becomes stronger  by changing the pertinent new free parameters.

In section 2  the generalized GS equation and the generalized Solov'ev solution are briefly reviewed and the method of solving the generalized GS equation adopting the generic linearized ansatz is presented.
Then, analytic solutions to the generic linearized generalized GS equation are constructed in section 3. In section 4 we examine how the Solov'ev configuration is modified by varying the additional free parameters associated with the pressure, the poloidal current  in conjunction with the parallel component of the velocity, and electric field, thus obtaining a variety of new configurations. The conclusions  are summarized in section 5.


\section{Generic linear axisymmetric  equilibria with non-parallel flow}

The generalized GS equation is given in SI units by (cf. \cite{tath,sim})
\be
 (1-M_p^2) \Delta^\star \psi -
         \frac{1}{2}\frac{d M_p^2}{d\psi} |\nabla \psi|^2
                     + \frac{1}{2}\frac{d}{d\psi}\left(\frac{X^2}{1-M_p^2}\right)
+ \mu_0 R^2 \frac {d P_s}{d\psi}+ \mu_0\frac{R^4}{2}\frac{d}{d\psi}\left[\frac{\rho}{1-M_p^2}
\left(\frac{d\Phi}{d\psi}\right)^2\right]
    = 0\,.
                            \label{orGS}
 \ee
 Here, $(z,R,\phi)$ are cylindrical coordinates with $z$
 corresponding to the axis of symmetry; 
 $\psi(R,z)$ is the poloidal magnetic flux function which labels the magnetic surfaces; $M_p(\psi)$ is
 the Mach function of the poloidal velocity with respect to the
 poloidal-magnetic-field Alfv\'en velocity;
 $\rho(\psi)$ and $\Phi(\psi)$ are  the density and  the electrostatic
 potential; $X(\psi)$ relates to the toroidal magnetic
 field and to the poloidal current (cf. relations (\ref{mf}) and (\ref{cd}));
 for vanishing flow the surface function $P_s(\psi)$
  coincides with the pressure (cf. (\ref{pr}));   $\Delta^\star=R^2\nabla\cdot(\nabla/R^2)$ and $\mu_0$ is the permeability of free space.

Equation (\ref{orGS}) can be  simplified  by  the  transformation
\begin{equation}
\label{tr}
u(\psi) = \int_{0}^{\psi}\left\lbrack 1 -
M^{2}(g)\right\rbrack^{1/2} dg,
\end{equation}
which reduces (\ref{orGS}) to 
\bea
\label{eq1}
\Delta^{*}u+\frac{1}{2}\frac{d}{du}\left[\frac{X^{2}}{1-M_{p}^{2}}\right]
+\mu_0 R^{2}\frac{dP_{s}}{du}+\mu_0 \frac{R^{4}}{2}\frac{d}{du}
\left[\rho\left(\frac{d\Phi}{du}\right)^{2}\right]=0\,. 
\eea
Note that no quadratic term as $|{\bf\nabla}u|^{2}$ appears
anymore in (\ref{eq1}). The transformation (\ref{tr}) does not affect the shape of magnetic surfaces; it just relabels them  in terms of the function $u$. It is noted here that there is a misprint of the respective $R^4$-term in Eq. (17) of \cite{sim}, i.e. the density factor should be linear instead of quadratic.  To solve Eq. (\ref{eq1}), the surface-quantity terms ($(1/2)(d/du)\left[X^{2}/(1-M_{p}^{2})\right]$, $\mu_0 R^{2}dP_{s}/du$  and $\mu_0 (R^{4}/2)(d/du)\left[\rho(d\Phi/du)^{2}\right]$) should be assigned as functions of $u$. Also, the equilibrium can be completely determined by calculations in the $u$-space; specifically, the magnetic field, velocity, current density, pressure and electric field are given by the relations:
\be
\label{mf}
\bB=I\nabla \phi +(1-M_p^2)^{-1/2}\nabla\phi\times \nabla u \,,
\ee
with 
\be 
\label{Iota}
 I=(1-M_p^2)^{-1/2}\left[\frac{X}{(1-M_p^2)^{1/2}}-\mu_0 R^2 \sqrt{\frac{\rho}{\mu_0}} M_p \frac{d \Phi}{du}\right]\,, 
\ee
\be
\label{vel}
\bv =\frac{M_p}{\sqrt{\mu_0 \rho}}\bB -R^2(1-M_p^2)^{-1/2}\frac{d\Phi}{du}\nabla \phi \,,
\ee
\be
\label{cd}
\bj=\frac{1}{\mu_0}\left\{(1-M_p)^{-1/2}\left[\Delta^\star u-\frac{1}{2(1-M_p^2)}\frac{d(1-M_p^2)}{du}|\nabla u|^2\right]\nabla \phi - \nabla \phi \times \nabla I\right\}\,,
\ee
\be
\label{pr}
P=P_s-\rho\left[\frac{v^2}{2}-R^2\left(\frac{d\Phi}{du}\right)^2\right]\,,
\ee
\be
\label{ep}
\bE=-\frac{d\Phi}{du}\nabla u 
\ee
with $v^2=\bv\cdot\bv$.

\begin{figure}[h]
\begin{center}
\includegraphics[width=0.5\linewidth]{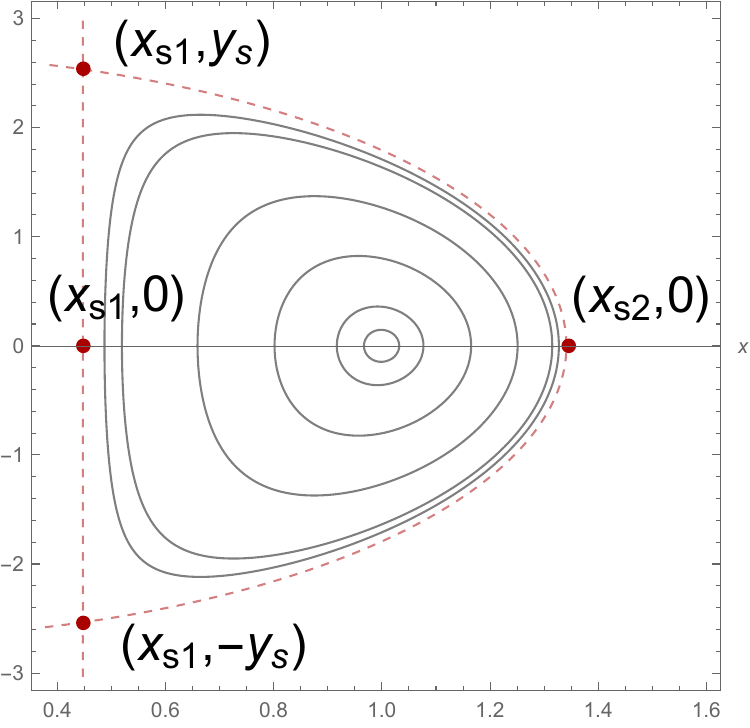}  
\end{center}
\caption{The equilibrium configuration of the GS solution (\ref{eq4}) on the poloidal plane $x-y$, with $x$ corresponds to the horizontal axis and $y$ to the vertical axis. The red-dashed curve represents the separatrix with  $x_{s1}=\sqrt{\epsilon}$,
 $x_{s2}=\left(-\epsilon \lambda -3\delta^2+\sqrt{3}\sqrt{-\epsilon^2\lambda^2-2 \delta^2\epsilon\lambda +3 \delta^4+\delta^2\lambda+4\lambda^2}/(2\lambda)\right)^{1/2}$ and $y_s=\left[(1-\epsilon)^2(2\delta^2+\lambda(\epsilon+1))\right]^{1/2}$.}
                                                          \label{Fig 0}
\end{figure}

Following Ref. \cite{sim} we first make the following choice of the free surface-function terms in (\ref{eq1})
\bea
\label{eq2}
\frac{d(\mu_0 P_{s})}{du}&=&-\frac{P_{s0}}{U_0}\,,\nn \\
\frac{1}{2}\frac{d}{du}\left[\frac{X^{2}}{1-M_{p}^{2}}\right]
&=&\epsilon\frac{P_{s0}R_{0}^{2}}{U_0(1+\delta^{2})}\,,\nn \\
\frac{1}{2}\frac{d}{du}\left[\mu_0 \rho\left(\frac{ d\Phi}{du}\right)^{2}\right]
&=&-\lambda\frac{P_{s0}}{U_0(1+\delta^{2})R_{0}^{2}}\,.
\eea
It is noted that a couple of the five surface quantities involved in the generalized GS equation remain free.
Normalizing as $x:=R/R_{0}$, $y:=z/R_{0}$, $U:=u/u_{0}$ with $u_{0}:=P_{s0}R_{0}^{4}/(2U_0(1+\delta^{2}))$, Eq.
(\ref{eq1}) assumes the form
\bea
\label{eq3}
U_{xx}-\frac{1}{x}U_{x}+U_{yy}-2\lambda x^{4}-2(1+\delta^2)x^{2}+2 \epsilon=0\,,
\eea
and the generalized Solov'ev solution is given by
\bea
\label{eq4}
U_{gs}=\left[y^{2}\left(x^{2}-\epsilon\right)+\frac{\delta^2+\lambda}{4}(x^{2}-1)^{2}+\frac{\lambda}{12}(x^{2}-1)^{3}\right]\,.
\eea
The equilibrium configuration for $\lambda >0$ is shown in Fig. \ref{Fig 0}.

As already mentioned in section 1,  the outermost closed surface (red-dashed line) is a separatrix consisting of an elliptic outer part and a straight inner part parallel to the axis of symmetry; thus, a couple of X-points are created. Expressions of the X-point coordinates and of the inner and outer points of the configuration on the mid-plane $y=0$ in terms of the parameters $\delta$, $\epsilon$ and $\lambda$ are given in the caption of Fig. \ref{Fig 0}. The magnetic axis is located at ($x=1,y=0$). The parameter $\epsilon$ determines the compactness of the configuration, i.e. for $\epsilon >0$ the equilibrium is diamagnetic and describes either a tokamak or a spherical tokamak, while for $\epsilon=0$ the inner part of the separatrix touches the axis of symmetry forming a spheromak. The parameter $\delta$ relates to the elongation of the magnetic surfaces in the vicinity of the magnetic axis, i.e. the higher the value of $\delta$, the larger the elongation parallel to the axis of symmetry. For $\lambda=0$ the velocity becomes parallel to the magnetic field, while for $\lambda <0$ a third X-point appears outside the separatrix. When  the velocity vanishes  ($M_p=\lambda=0$),  the usual static Solov'ev solution is recovered. Here we will consider equilibria with $\lambda >0$.

We now try the following new generic linearizing ansatz to Eq. (\ref{eq1})
\bea
\label{eq5}
\frac{d(\mu_0 P_{s})}{du}&=&-\frac{P_{s0}}{U_0}+\bar{\alpha} \frac{P_{s0}}{U_0^{2}}u\,,\nn
\\
\frac{1}{2}\frac{d}{du}\left[\frac{\mu_0 X^{2}}{1-M_{p}^{2}}\right]&=&
\epsilon\frac{P_{s0}R_{0}^{2}}{U_0(1+\delta^{2})}-\bar{\beta}  
\epsilon\frac{P_{s0}R_{0}^{2}}{U_0^{2}(1+\delta^{2})}u\,,\nn
\\
\frac{1}{2}\frac{d}{du}\left[\mu_0 \rho\left(\frac{d\Phi}{du}\right)^{2}\right]
&=&-\lambda\frac{P_{s0}}{U_0(1+\delta^{2})R_{0}^{2}}+
\bar{\gamma}   \frac{\lambda P_{s0}}{U_0^{2}(1+\delta^{2})R_{0}^{2}}u\,,
\eea
where $\bar{\alpha}, \bar{\beta}, \bar{\gamma}   $ are dimensionless parameters. 
The linear  terms in $u$  in (\ref{eq5}) result in more realistic equilibrium characteristics,  e.g. the constant $\epsilon \left[P_{s0}R_{0}^{2}/(U_0(1+\delta^{2}))\right]$ in (\ref{eq2}) results in monotonically increasing toroidal-current-density profiles while the additional  linear in $u$ term, $-\bar{\beta}  
\epsilon \left[P_{s0}R_{0}^{2}/(U_0^{2}(1+\delta^{2})\right]u$, in  (\ref{eq5}) results in peaked toroidal-current-density profiles. Also, the linear term $\bar{\gamma} \left[\lambda P_{s0}/(U_0^{2}(1+\delta^{2}R_{0}^2))\right] u$ in (\ref{eq5}) produces larger velocity shear.
Setting  $G:=2\epsilon$, $F:=-2(1+\delta^{2})$, $E:=-2\lambda$,
$\alpha :=\bar{\alpha}  (u_{0}/U_0)$, $\beta  :=\bar{\beta} (u_{0}/U_0)$
and $\gamma   :=\bar{\gamma}    (u_{0}/U_0)$ as well as substituting (\ref{eq5}) into (\ref{eq1}) we straightforwardly obtain the following equation:
\bea
\label{eq6}
U_{xx}-\frac{1}{x}U_{x}+U_{yy}+Ex^{4}+Fx^{2}+G-Ex^{4}\gamma    U
-Fx^{2}\alpha  U-G\beta   U=0\,.
\eea
Note that the solution to (\ref{eq6})  for $\alpha =\beta  =\gamma   =0$ is  given by (\ref{eq4}).
We now will pursue a solution to (\ref{eq6}) in the form of an expansion about the GS solution (\ref{eq4}), i.e.
\be
\label{eq7}
U=U_{gs}+W(x,y)\,,
\ee
with the function $W(x,y)$ to be determined so that (\ref{eq7}) to satisfy (\ref{eq6}).
Substituting (\ref{eq7}) into (\ref{eq6}) yields
\bea
\label{eq8}
W_{xx}-\frac{1}{x}W_{x}+W_{yy}-\gamma    Ex^{4}W-\alpha  Fx^{2}W
-\beta  GW=\nn \\=(\gamma   Ex^{4}+\alpha Fx^{2}+\beta  G)U_{gs}\,.
\eea
This is an inhomogeneous linear second-order PDE
for the unknown function $W(x,y)$. The general solution
consists of the sum of the general solution to the counterpart homogeneous  equation plus
a particular solution of the complete inhomogeneous equation.


\section{Solutions}\

We first consider  the homogeneous counterpart of (\ref{eq8}):
\bea
\label{eq12}
W_{xx}-\frac{1}{x}W_{x}+W_{yy}-(\gamma   Ex^{4}+\alpha Fx^{2}+\beta  G)W=0\,.
\eea
The general solution of (\ref{eq12}) can be found by separation of variables, i.e. in the form
\bea
\label{eq13}
W(x,y)=X(x)Y(y)\,.
\eea
Inserting (\ref{eq13}) into (\ref{eq12}) we find:
\bea
\label{eq14}
Y^{''}+k^{2}Y=0\,,
\eea
and
\bea
\label{eq15}
X^{''}-\frac{1}{x}X^{'}-(\gamma   Ex^{4}+\alpha Fx^{2}+\beta  G+k^{2})X=0\,,
\eea
where $k$ is the parameter of separation.
The general solution of  (\ref{eq14}) is
\bea
\label{eq23}
Y(y)=C\cos(ky)+D\sin(ky)\,.
\eea
Setting $\ksi=x^{2}$,  Eq. (\ref{eq15}) takes the form
\bea
\label{eq15a}
4\ksi X^{''}(\ksi)-(\gamma   E\ksi^{2}+\alpha F\ksi+\beta  G +k^2) X=0\,.
\eea
Since $\xi=0$ is a regular singular point of (\ref{eq15a}) its solution can be derived by the method of Frobenius in terms of  infinite converging series of the form:
\bea
\label{eq24}
X(\ksi)=\sum_{n=0}^{\infty}a_{n}\ksi^{n+r}\,.
\eea
Substituting (\ref{eq24}) into (\ref{eq15a})
results in the following recursion relations:
\bea
\label{eq25}
4r(r-1)a_{0}=0\,,\nn \\
4r(r+1)a_{1}-(\beta   G+k^{2})a_{0}=0\,,\nn \\
4(r+1)(r+2)a_{2}-\alpha Fa_{0}-(\beta   G+k^{2})a_{1}=0\,,\nn \\
(n\geq 2)\; \; \; 4(n+r)(n+r+1)a_{n+1}-\gamma   Ea_{n-2}-\alpha Fa_{n-1}-(\beta   G+k^{2})a_{n}=0\,,
\eea
where $a_{0}$ is a freely specified parameter. The roots of the indicial polynomial, $r(r-1)$, are $r_1=1$ and $r_2=0$. Therefore, for $r_1=1$,  one  solution of (\ref{eq15a}) is
\be
\label{eq25a1}
X_1(\ksi)=\ksi\sum_{n=0}^{\infty}a_{n}\ksi^n\,.
\ee
Since $r_1-r_2$ is a non-zero integer, the second linearly independent solution is of the form
\be
\label{eq25a2}
X_2(\ksi)=c X_1(\ksi)\ln \ksi + \sum_{n=0}^{\infty}\bar{a}_{n}\ksi^n\,,
\ee
where the coefficients $c$ and $\bar{a}_n$ can be determined by substituting (\ref{eq25a2}) into (\ref{eq15a}).
Here, we will employ the solution $X_1(\ksi)$.
By means of (\ref{eq13}), (\ref{eq23}), (\ref{eq24}) and introducing a new parameter $s$ by $k=s l$ with $l=0,1,2,\ldots$, the solution of the homogeneous PDE (\ref{eq12}) is written as:
\bea
\label{gsol}
W(x,y)=\sum^\infty_l X_1(\ksi)
\left[C_l \cos (s l y) + D_l \sin(s l y)\right] \nn \\
=\sum_{l=0}^\infty \sum_{n=0}^{\infty} a_{n,sl}\ksi^n\left[C_l \cos(s l y)+ D_l \sin(s l y)\right]\,. & & 
\eea 
The values of the free coefficients $C_l$ and $D_l$ may be properly fixed in order to impose a boundary. However, here since in the Solov'ev equilibrium 
the boundary (separatrix) is spontaneously formed we will not impose any boundary condition; accordingly, in (\ref{gsol}) we will keep  only the  first two $l$-terms (for $l=0$ and $1$).

Furthermore, we will pursue a particular solution of (\ref{eq8}) in the form:
\bea
\label{eq9}
W(x,y)=W_{1}(x)y^{2}+W_{2}(x)\,.
\eea
Inserting (\ref{eq9}) into (\ref{eq8}) yields
\bea
\label{eq10}
W_{1}^{''}-\frac{1}{x}W_{1}^{'}-(\gamma   Ex^{4}+\alpha Fx^{2}+\beta  G)W_{1}
=\left(x^{2}-\frac{G}{2}\right)(\gamma   Ex^{4}+\alpha Fx^{2}+\beta  G)\,,
\eea
and
\bea
\label{eq11}
W_{2}^{''}-\frac{1}{x}W_{2}^{'}+2W_{1}-(\gamma   Ex^{4}+\alpha Fx^{2}+\beta  G)W_{2}
=\nn \\=-
\left[\frac{(F+E+2)}{8}(x^{2}-1)^{2}+\frac{E}{24}(x^{2}-1)^{3}\right]
(\gamma   Ex^{4}+\alpha Fx^{2}+\beta  G)\,.
\eea
Equation (\ref{eq10}) involves only the function $W_1(x,y)$. Thus, once (\ref{eq10}) is solved, the solution contributes to the inhomogeneous part of (\ref{eq11}).

To solve (\ref{eq10}) we again change variable, $\ksi=x^{2}$, to get:
\bea
\label{eq16}
4\ksi W_{1}^{''}(\ksi)-(\gamma   E\ksi^{2}+\alpha F\ksi +\beta  G)W_{1}=
\left(\ksi-\frac{G}{2}\right)(\gamma   E\ksi^{2}+\alpha F\ksi+\beta  G)\,.
\eea
Setting
\bea
\label{eq17}
W_{1}+\left(\ksi-\frac{G}{2}\right):=\bar{W}_{1}\,,
\eea
Equation (\ref{eq16}) is put in the form:
\be
\label{eq17a}
4\ksi\bar{W}_1^{\prime\prime}-(\gamma   E\ksi^{2}+\alpha F\ksi+\beta  G)\bar{W}_1=0\,.
\ee
For $\beta  =0$, the solution of (\ref{eq17a}) is:
\be
\bar{W}_1=c_1 Ai\left(\frac{\alpha  F+\gamma   E \ksi}{(2\gamma    E)^{2/3}}\right)
         +c_2 Bi\left(\frac{\alpha  F+\gamma   E \ksi}{(2\gamma    E)^{2/3}}\right)\,,
\ee
where $Ai$ and $Bi$ are the Airy functions of the first and second kind.
For $\beta  \neq 0$ apart from the $k^2$-term, Eq. (\ref{eq15a}) is identical to (\ref{eq17a}). Therefore the solution of (\ref{eq17a}) is already included in (\ref{gsol}) for $l=0$.

Finally, we will obtain a particular solution of the inhomogeneous equation ({\ref{eq11}).  Introducing once more the variable  $\ksi=x^{2}$,
Eq. (\ref{eq11}) is written as:
\bea
\label{eq19}
4\ksi W_{2}^{''}(\ksi)-(\gamma   E\ksi^{2}+\alpha F\ksi+\beta  G)W_{2}=
\nn \\=-2W_{1}(\ksi)-[C_{5}\ksi^{5}+C_{4}\ksi^{4}+C_{3}\ksi^{3}+C_{2}\ksi^{2}+C_{1}\ksi+C_{0}]\,,
\eea
where $W_1=\bar{W}_1-(\ksi-G/2)$ by  (\ref{eq17}) and
\bea
\label{eq22}
C_{5}&=&\gamma   \frac{E^{2}}{24}\,,\nn \\
C_{4}&=&\frac{\gamma   E(F+2)}{8}+\alpha F\frac{E}{24}\,,\nn
\\
C_{3}&=&\beta  G\frac{E}{24}+\alpha F\frac{(F+2)}{8}-\gamma   E\frac{(2F+E+4)}{8}\,,\nn
\\
C_{2}&=&\beta  G\frac{(F+2)}{8}-\alpha F\frac{(2F+E+4)}{8}+\gamma   E\frac{(3F+2E+6)}{24}\,,\nn
\\
C_{1}&=&-\beta  G\frac{(2F+E+4)}{8}+\alpha F\frac{(3F+2E+6)}{24}\,,\nn
\\
C_{0}&=&\beta  G\frac{(3F+2E+6)}{24}\,.
\eea
In view of the converging series solutions (\ref{eq25a1})  to the homogeneous equations (\ref{eq15a}) and (\ref{eq17a}), we pursue  a similarly structured solution  to ({\ref{eq19}):
\bea
\label{eq20}
W_{2}=\ksi\sum_{n=0}^{\infty}b_{n}\ksi^n\,.
\eea
Inserting (\ref{eq20}) into (\ref{eq19}) we obtain after some tedious but straightforward  algebra that $b_{0}$ is freely specified and
\bea
\label{eq21}
C_{0}+G=0\,,\nn \\
8b_{1}-\beta  Gb_{0}+2d_{0}-2+C_{1}=0\,,\nn \\
24b_{2}-\alpha Fb_{0}-\beta  Gb_{1}+2d_{1}+C_{2}=0\,,\nn \\
48b_{3}-\gamma   Eb_{0}-\alpha Fb_{1}-\beta  Gb_{2}+2d_{2}+C_{3}=0\,,\nn \\
80b_{4}-\gamma   Eb_{1}-\alpha Fb_{2}-\beta  Gb_{3}+2d_{3}+C_{4}=0\,,\nn \\
120b_{5}-\gamma   Eb_{2}-\alpha Fb_{3}-\beta  Gb_{4}+2d_{4}+C_{5}=0\,,\nn \\
(n\geq 6)\; \; \; 4n(n+1)b_{n}-\gamma   Eb_{n-3}-\alpha Fb_{n-2}-\beta  Gb_{n-1}+2d_{n-1}=0\,.
\eea
Inspection of the last of  Eqs. (\ref{eq22}) and $C_0+G=0$ (first of Eqs. (\ref{eq21})) together with the definitions $G:=2\epsilon$, $F:=-2(1+\delta^{2})$ and $E:=-2\lambda$ implies the following:

\begin{enumerate}
\item 
When $\beta =0$, then $\epsilon=0$   and therefore the second of (\ref{eq5})
becomes \newline $(1/2)d/du(X^2/(1-M_p^2))=0$. In the absence of flow then (\ref{Iota}) implies that the plasma is
confined by a vacuum toroidal magnetic field ($\beta  _p=1$ equilibrium); when the non-parallel
flow is there the electric-field term in (\ref{Iota}) additionally contributes to the poloidal current. 
\item When $\epsilon\neq 0$  and $\beta  \neq 0$, the  last of (\ref{eq22}) becomes  $\beta  (3F+2E+6)+24=0$. In this case  all the linear terms in the ansatz (\ref{eq5}) can remain finite. This
relation just acts as a constraint on the parameters $\beta$, $\delta$ and $\lambda$.  In section 4 we intend  to examine the impact of the linear in $u$ electric-field term in (\ref{eq5}) on the equilibrium  by varying the parameter $\beta$; accordingly,  leaving $\beta$ free we will employ the constraint to express $\delta$  in terms of $\beta$  and $\lambda$, i.e.
\be  
\delta=\left[\frac{(2/3)(6-\beta   \lambda)}{\beta  }\right]^{1/2} \,.
                                                    \label{delta}
\ee
\item When $\epsilon=0$, $\beta$ does not necessarily vanish and both $\delta$ and $\lambda$ remain free.
\end{enumerate}
%
%
Additional free parameters are $\alpha $, $\beta  $, $\gamma$, $a_0$,  $b_{0}$, $C_0$, $C_1$, $D_1$, $R_0$, $P_{s0}$, $U_0$ and $s$.
It is recalled here that  $R_0$ is a  reference length determining the $R$-position of the magnetic axis; the parameters $P_{s0}$ and $U_0$ are  reference  parameters associated with the pressure and poloidal magnetic flux;   $\delta$ and $\epsilon$ are geometrical (shaping) parameters of the starting generalized Solov'ev equilibrium (\ref{eq4}) shown in Fig. \ref{Fig 0}; $\lambda$ relates to the electric field term in (\ref{eq4}) through the ansatz (\ref{eq2}); $\alpha$, $\beta$ and $\gamma$ are physical parameters of the current density, pressure and electric field, respectively, in connection to the additional linear terms in the ansatz (\ref{eq5}); and $a_0$, $b_0$, $C_0$, $C_1$,   $D_1$ and $s$ are shaping parameters associated with the new constructed solution. A list of the free parameters is given in Table \ref{table}. Also, it is noted that the series expansion (\ref{eq20})
was shown numerically to  converge and we kept terms up to the order of $\ksi^{20}$ in our calculations.

\begin{table}[]
\caption{\bf List of the free parameters}
\label{table}
\begin{tabular}{|l|l|}
\hline
{\em Parameters} & {\em Description }  \\  \hline\hline
$R_0$    & Reference $R$-position of the magnetic axis \\ \hline
$P_{s0}$, $U_0$ &  Reference pressure and poloidal magnetic flux  \\ \hline
$\delta$, $\epsilon$ & Shaping of the generalized Solov'ev equilibrium of Fig. (\ref{Fig 0})   \\ \hline
$\lambda$   & Parameter related to the electric field term in the \\
     &   generalized Solov'ev solution (\ref{eq4})  \\ \hline
$\alpha$, $\beta$, $\gamma$ & Current density, pressure and electric field parameters \\
&   in the ansatz (\ref{eq5})   \\ \hline
$a_0$, $b_0$, $C_0$, $C_1$, $D_1$, $s$  & Shaping parameters associated  with the new constructed \\
& solution  \\ \hline
$\rho_0$, $l_\rho$, $M_{p0}$, $l_{M_p}$  & Density and poloidal Mach function parameters (Eq. (\ref{choice})) \\ \hline
\end{tabular}
\end{table}

In summary, a solution of the general linearized form of the generalized GS equation (\ref{eq8})
consists of a separated solution of its homogeneous counterpart  in
terms of (\ref{eq13}),  (\ref{eq23}) and (\ref{eq25a1}) plus a particular
solution of the form (\ref{eq9}), expressed in terms of the series (\ref{gsol}) for $l=0$ and (\ref{eq20}).
\begin{figure}[h]
\begin{center}
\includegraphics[width=0.45\linewidth]{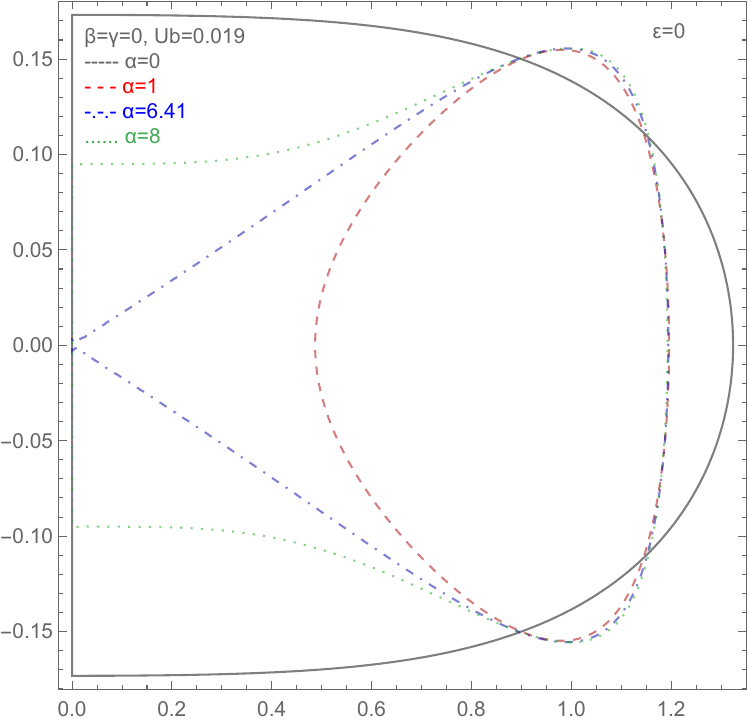} \hspace{0.5cm}
\includegraphics[width=0.45\linewidth]{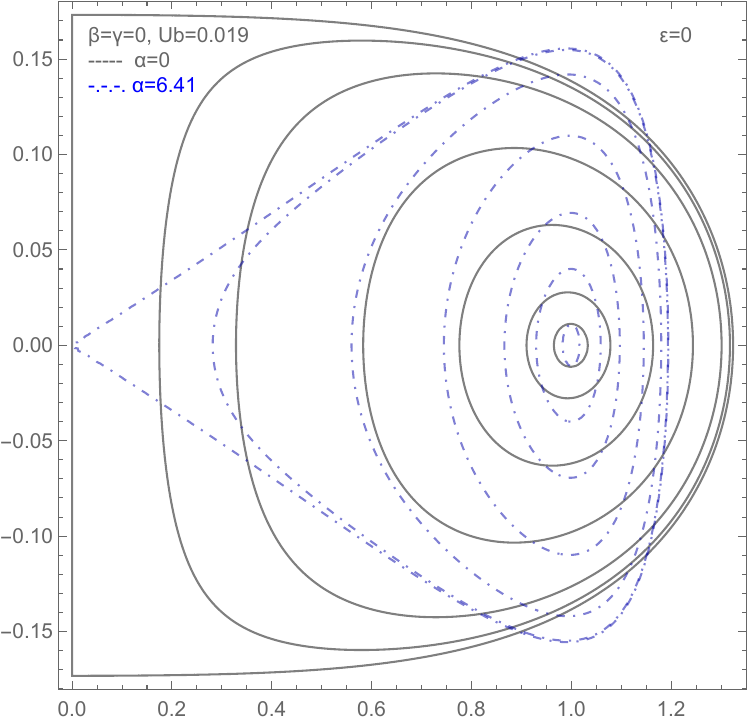}
\end{center}
\caption{Left: different equilibrium boundaries are given on the poloidal plane $x-y$ as the additional linear-in-$U$ pressure term in the first of (\ref{eq5}) varies  in connection with different values of $\alpha$. The black continuous curve represents the separatrix of the respective genaralized Solov'ev spheromak configuration. Right: a set of magnetic surfaces for one of the configuration representd by blue, dot-dashed U-curves are shown together with the respective curves of the genaralized Solov'ev equilibrium.  $U_b$ is the value of the bounding $u$-curve (separatrix).}

                                                          \label{Fig 1}
\end{figure}

\section{Equilibrium configurations} 
By employing the solution to (\ref{eq6}) derived  in section 3
 we have 
constructed several  new equilibria. Particularly, we  examined
how the generalized Solov'ev equilibrium is modified by varying the  additional
parameters $\alpha $,  $\beta$ and $\gamma$, associated with the
respective linear-in-$u$ terms in (\ref{eq5}) of pressure, poloidal-current in conjunction with the parallel velocity (Eqs. (\ref{Iota}) and (\ref{cd})), and electric field. To impose a stagnation point, $(x_0,y_0)$, in addition to the conditions $dU/dx|_{x_0,y_0}=0$ and $dU/dy|_{x_0,y_0}=0$ we considered the sign of the quantity  
$$\left.\frac{d^2 U}{dx^2}\frac{d^2 U}{dy^2}-\left(\frac{d^2 U}{dx dy}\right)^2\right|_{(x_0,y_0)}\,,$$ which should be positive for a magnetic axis and negative for an X-point.
For the up-down symmetric configurations considered,  the magnetic axis is located  at the same position, $(x=1, y=0)$ as the generalized  Solov'ev configuration and  $U=0$ thereon. Note that in this case the condition $dU/dy|_{(1,0)}=0$ is identically satisfied because of symmetry.
 If not otherwise specified,  we assigned the following parametric values: $s=0.1$, $\delta=0.1$, $\lambda=0.1$ and $a_0=1$. It is recalled that when $\beta\neq 0$ and $\epsilon \neq 0$, $\delta$ should be calculated from (\ref{delta}). The parameters $C_0$, $C_1$ and $D_1$ were determined as solutions to the set of the algebraic equations derived by the  conditions associated with the imposition of a stagnation point.  Since we employed dimensionless quantities, the reference dimensional parameters $R_0$, $P_{s0}$ (in units of $4 \pi \times 10^{-7}$ H/mPA ) and $U_0$ remained free. 
 Depending on the  value of $\epsilon$ we constructed tokamak, spherical tokamak and spheromak equilibria, the latter corresponding to $\epsilon=0$.
 
 First, we examined the impact of the $\alpha$- $\beta$- and $\gamma$-terms individually on up-down symmetric equilibria  by imposing the same position of the magnetic axis, $(x=1,y=0)$,  as that of the respective generalized Solov'ev equilibrium and the  same value $U=0$ thereon. In the presence of $\alpha$-term, a D-shaped configuration is formed with negative triangularity,  the boundary of which is shown in Fig. \ref{Fig 1}-left by the red-dashed curve. As $\alpha$ gradually takes larger values the configuration with increasing triangularity extends toward the axis of symmetry, eventually touching it  to form a separatrix with a single X-point located at the origin of the coordinate system  (blue, dot-dashed line) and then becomes a peculiar configuration having a separatrix with a couple of X-points (green-dotted line).
 
 The $\beta$-term creates configurations with small triangularity elongated along the $x$-axis as shown in Fig. \ref{Fig 2}. The red-dashed curves represent the magnetic surfaces of a configuration with the same poloidal magnetic flux as the respective generalized Solov'ev tokamak configuration represented by the black-continuous curve. As $\beta$ takes larger values the configuration  extends further  and eventually reaches the axis of symmetry transforming into  a spheromak. It is noted that for $\epsilon=\beta=\gamma=0$ the equilibrium becomes  independent of $\beta$.
The impact of the $\gamma$-term is similar to that of the $\alpha$-term as  shown in Fig. \ref{Fig 3}.

Subsequently, the impact of combinations of the $\alpha$-, $\beta$- and $\gamma$-terms was examined by imposing an X-point at several positions. Thus, a variety of configurations were created. Three examples are given in Figs. \ref{Fig 4}-\ref{Fig 6}.  In Fig. \ref{Fig 4} the blue, dot-dashed curves represent an up-down symmetric  spherical tokamak configuration ($\epsilon=0.1$) the X-points of which are common with those of the respective 
  generalized Solov'ev equilibrium, but the poloidal magnetic flux of this former equilibrium is larger than that of the latter. The red-dashed curve represents the boundary of a D-shaped configuration with poloidal magnetic flux equal to that of the generalized Solov'ev  solution. In Fig. \ref{Fig 5} is shown an up-down symmetric diverted configuration with the imposed X-points located at ($x=0.84, y=\pm 14.5$), while  Fig. \ref{Fig 6} shows an up-down asymmetric configuration  with the magnetic axis located at ($x=1.02,y=0.83$) and the single lower X-point  at $(x=0.94, y=-22.6)$. 

\begin{figure}[h]
\begin{center}
\includegraphics[width=0.45\linewidth]{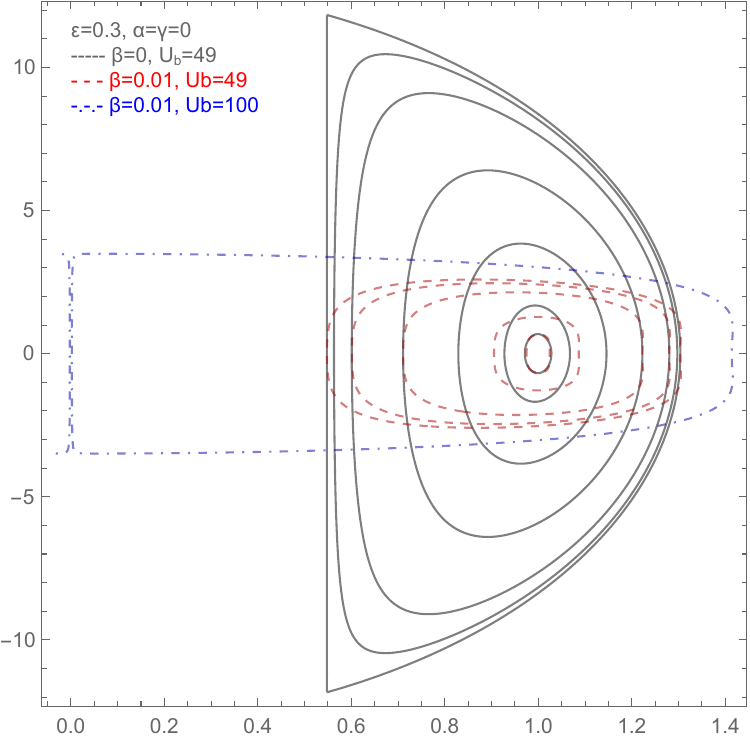} 
\end{center}
\caption{Fig. \ref{Fig 2} shows the impact of  the additional linear-in-$U$, poloidal-current  term in the second of (\ref{eq5}) in connection with different values of $\beta$. The black continuous curves represent the magnetic surfaces of the respective generalized Solov'ev tokamak configuration.  $U_b$ is the value of the bounding $u$-curve (separatrix).}

                                                          \label{Fig 2}
\end{figure}
\begin{figure}[h]
\begin{center}
\includegraphics[width=0.45\linewidth]{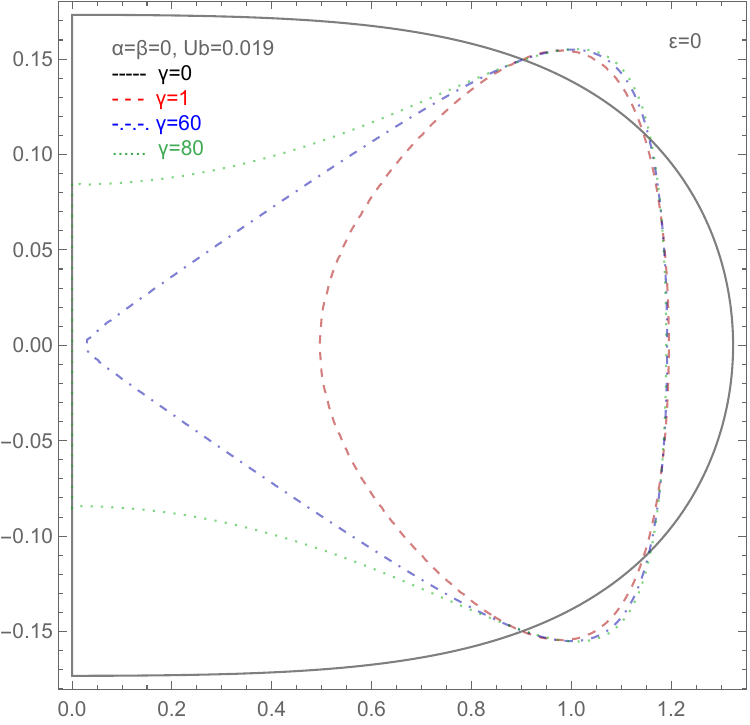} \hspace{0.5cm}
\includegraphics[width=0.45\linewidth]{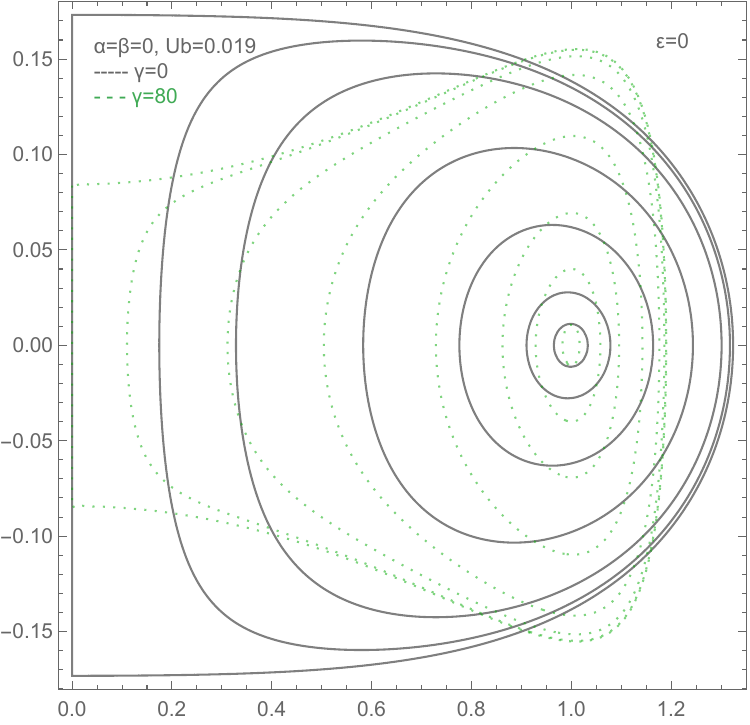}
\end{center}
\caption{Left: different equilibrium boundaries are given as the additional linear-in-$U$, electric-field term in the third of (\ref{eq5}) varies in connection with different values of $\gamma$. The black continuous curve represents the separatrix of the respective generalized Solov'ev spheromak configuration.  Right: A set of magnetic surfaces for one of the configuration represented by green-dotted U-curves are shown together with the respective curves of the Solov'ev equilibrium.  $U_b$ is the value of the bounding $u$-curve (separatrix).}

                                                          \label{Fig 3}
\end{figure}

\begin{figure}[h]
\begin{center}
\includegraphics[width=0.5\linewidth]{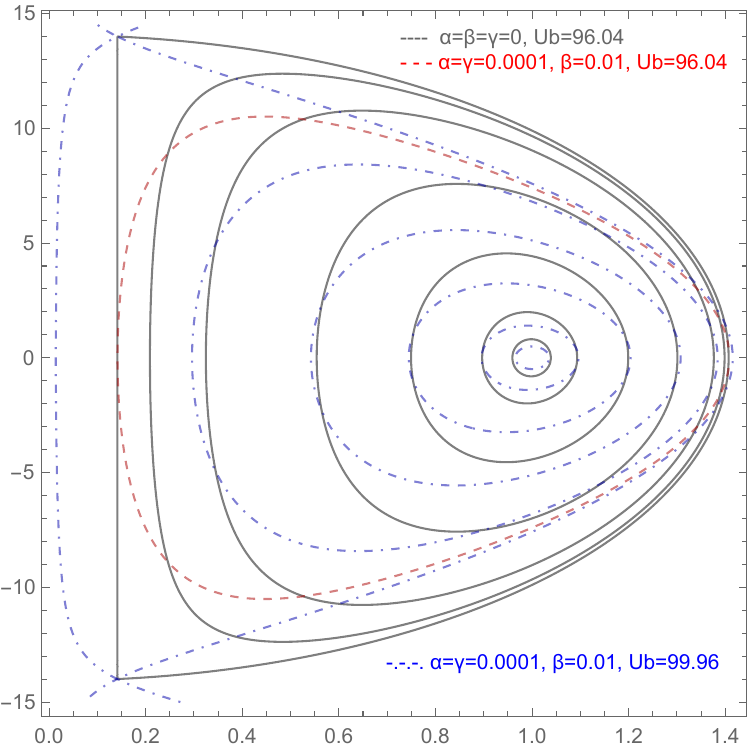} 
\end{center}
\caption{A spherical tokamak diverted configuration ($\epsilon=0.1$, $\delta=20$) with a couple of X-points common with the respective Solov'ev equilibrium. The red-dashed curve represents the boundary of a D-shaped tokamak equilibrium with poloidal magnetic flux equal to that of the Solov'ev one.  $U_b$ is the value of the bounding $u$-curve (separatrix).}

                                                          \label{Fig 4}
\end{figure}
\begin{figure}[h]
\begin{center}
\includegraphics[width=0.5\linewidth]{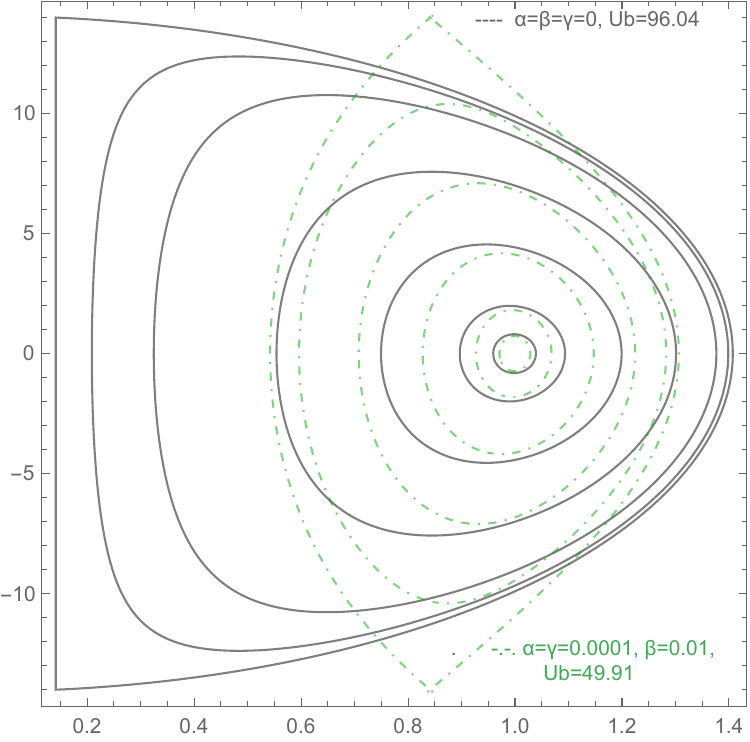} 
\end{center}
\caption{An  up-down symmetric, diverted tokamak configuration ($\epsilon=0.1,\ \delta=20$) with a couple of X-points located at ($x=0.84, y=\pm 14.5$).  The black continuous curves represent the magnetic surfaces of  the respective Solov'ev equilibrium.  $U_b$ is the value of the bounding $u$-curve (separatrix).}

                                                          \label{Fig 5}
\end{figure}
\begin{figure}[h]
\begin{center}
\includegraphics[width=0.5\linewidth]{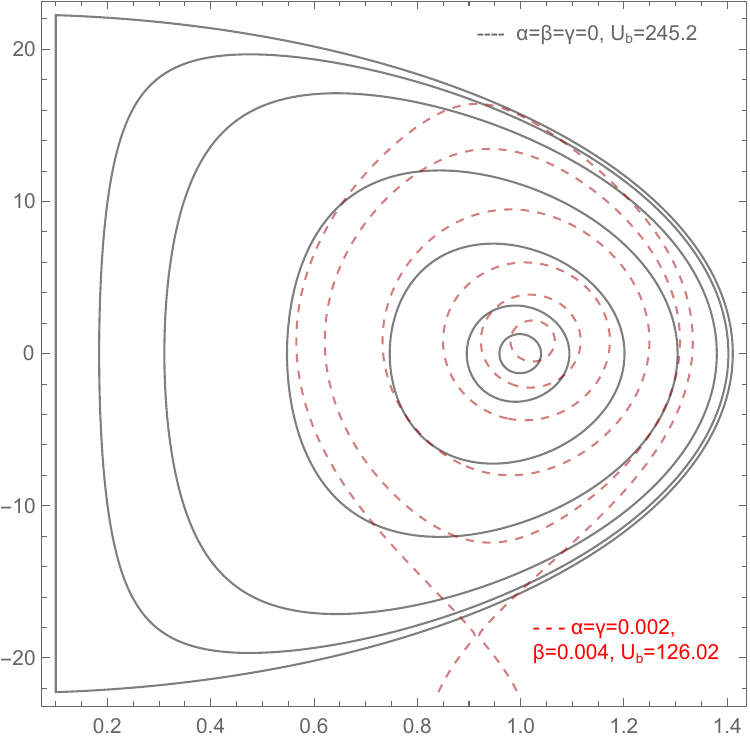} 
\end{center}
\caption{An  up-down asymmetric, diverted tokamak configuration for $k=0.2$, $\epsilon=0.2$, $\delta=31.6 $, $D_1=0.5$ and $D_2=0.01$. The magnetic axis is located at $(x=1.02, y=0.83)$ and the single lower X-point at $(x=0.94, y=-22.6)$. The black continuous curves represent the magnetic surfaces of  the respective Solov'ev equilibrium.  $U_b$ is the value of the bounding $U$-curve (separatrix).}

                                                          \label{Fig 6}
\end{figure}

Finally, to completely determine the equilibrium two from the five free functions involved should be specified. For example, considering as free functions the density 
 and the poloidal Mach function  we make the choice
\be
\label{choice}
\rho=\rho_0 \left(1-\frac{U}{U_b}\right)^{l_\rho}\,, \ \ 
M_p=M_{p0}\left(1-\frac{U}{U_b}\right)^{l_{M_p}}\,,
\ee
where $\rho_0$, $M_{p0}$, $l_\rho$ and $l_{M_p}$ are free parameters. Introducing the dimensionless quantities 
$$
\tilde{\rho}=\frac{\rho}{\rho_0}\,, \ \ \tilde{X}=\frac{X}{u_0/R_0}\,, \ \  \tilde{\bB}=\frac{\bB}{u_0/R_0^2}\,, \ \  \tilde{\bj}=\frac{\bj}{u_0/(R_0^3\mu_0)}\,, \ \  \tilde{\bv}=\frac{\bv}{u_0/(R_0^2 \sqrt{\mu_0 \rho_0})}\,, 
$$
$$   \tilde{\Phi}=\frac{\Phi}{u_0^2/(R_0^3\sqrt{\mu_0 \rho_0})}\,, \ \ \tilde{P}=\frac{P}{P_{s0}/(u_0^2/R_0^4)}\,, $$ 
integration of the relations (\ref{eq5}) with respect to $u$ yields (in the dimensionless form and omitting  the tildes)
\bea
\label{integr}
P_s&=& A-2(1+\delta^2)U+\alpha(1+\delta^2)U^2 \nn \\
\frac{X^2}{1-M_p^2}&=&B+4\epsilon U-2 \beta U^2  \nn \\
\rho\left(\frac{d\Phi}{dU}\right)^2&=&C-4 \lambda U +2 \gamma U^2\,,
\eea
where $A$, $B$ and $C$ are free parameters. The reference density $\rho_0$ remains free.
Using (\ref{integr}) in (\ref{pr}) we obtain
\be
\label{pr1}
P=A-2(1+\delta^2)U+\alpha(1+\delta^2)U^2- \frac{1}{2}\rho v^2+x^2\left(C-4 \lambda U +2 \gamma U^2\right).
\ee
The values of $A$ and $C$ are fixed so that the pressure vanish on the boundary.
Isobaric contours for the D-shaped tokamak configuration of Fig. \ref{Fig 4}, the boundary of which is indicated by the red-dotted curve therein,  are given in Fig. \ref{Fig 7} together with pressure profiles on the poloidal plane at $y=0$ and at the vertical line passing from the position of the pressure maximum. It is noted that although the values of the dimensionless pressure are quite high, realistic pressure values can be obtained by appropriate values of the free parameters $P_{s0}$ and $U_0$. For flows of experimental fusion pertinence the deviation of the magnetic surfaces from the isobaric surfaces is very small.

\begin{figure}[h]
\begin{center}
\includegraphics[width=0.5\linewidth]{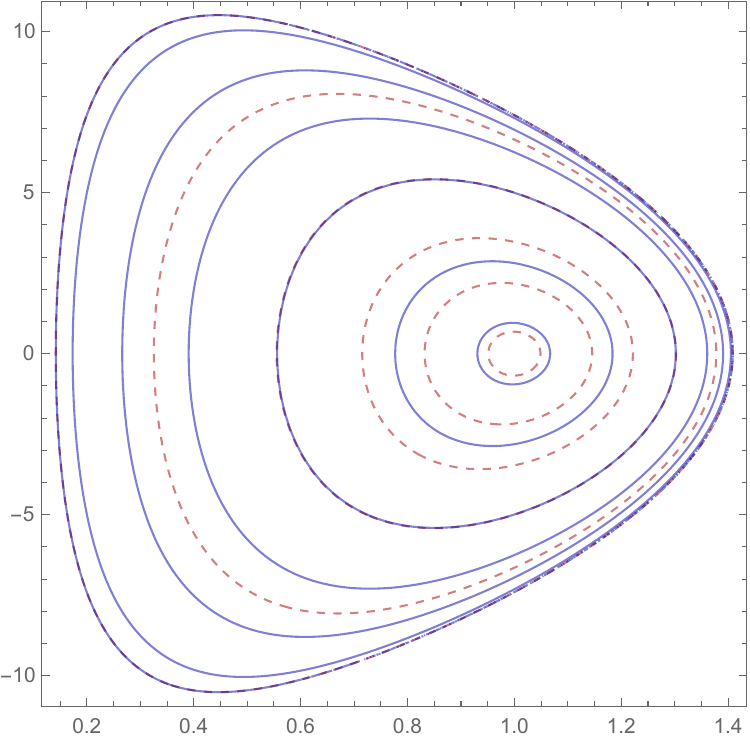}\newline
\vspace{1cm}
\includegraphics[width=0.45\linewidth]{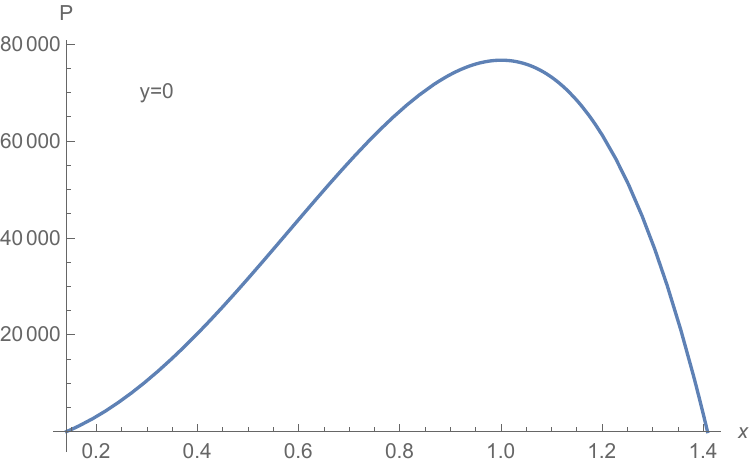} \hspace{0.5cm}
\includegraphics[width=0.45\linewidth]{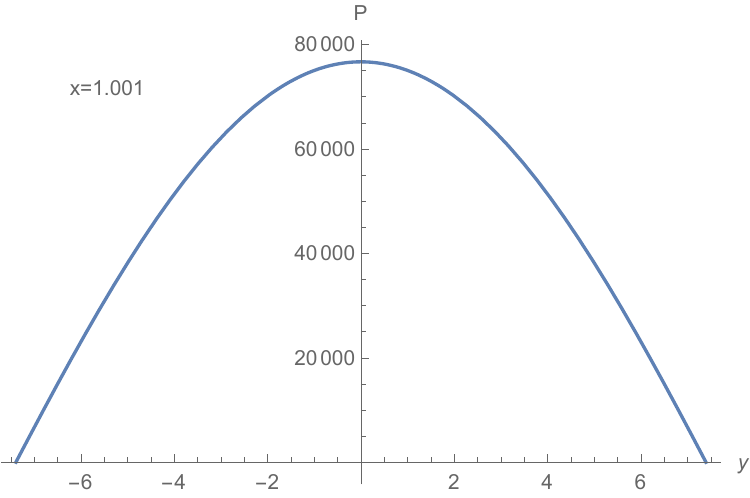}
\end{center}
\caption{Upper: Isobaric (blue-continuous) curves on the poloidal plane for the D-shaped tokamak configuration of Fig. (\ref{Fig 4}) for $B=0.2$, $M_{p0}=0.1$, $l_\rho=2$ and $l_{M_p}=2$. The red-dashed curves represent magnetic surfaces. Lower left: Respective pressure profile on the mid-plane $y=0$. Lower right: Respective vertical pressure profile at the position of the pressure maximum}

                                                          \label{Fig 7}
\end{figure}

\section{Conclusions}

Making a generic linearizing choice of the free function-terms involved in the generalized GS equation (\ref{eq1}) we solved the resulting linearized equation (\ref{eq8}) analytically. The requested solution was expressed as an expansion of the generalized Solov'ev solution (\ref{eq4}) in the form (\ref{eq7}) involving the determinable function $W$. Then, the solution was  obtained as a superposition  of the counterpart homogeneous equation and a particular solution of the complete inhomogeneous equation in  the form (\ref{eq9}). The solutions of the pertinent radial ODEs were expressed in terms of infinite converging series.

Employing these solutions, we constructed several configurations by changing the values of the free parameters, either individually or in combination,  associated with the new terms of pressure, poloidal current in conjunction with the parallel velocity, and electric field. The equilbria constructed are pertinent  to tokamaks, spherical tokamaks and spheromaks; they  include D-shaped configurations with either positive or negative triangularity and diverted configurations  with either a couple of X-points or a single X-point.

The present study can be extended to a more generic generalized GS equation accounting for CGL pressure anisotropy (\cite{evth}, Eq. (34) therein). Another extension could be pursued in the framework of Hall-MHD, a simplified two-fluid model with  inertialess electron-fluid elements keeping on magnetic surfaces and ion-fluid elements departing from them. 

\clearpage
\newpage

\section*{Aknowledgments}\

This work was conducted in the framework of participation of the University of Ioannina in the National Programme for the Controlled
Thermonuclear Fusion, Hellenic Republic.
The authors would like to thank the anonymous reviewers for constructive comments, 
which have resulted in a remarkably improved version of the manuscript.



\begin{thebibliography}{99}

\bibitem{HELENA1} C. Konz, W. Zwingmann, F. Osmanlic, B. Guillerminet, F. Imbeaux, P. Huynh, M. Plociennik, M. Owsiak, T. Zok, and M. Dunne, ``\textit{First physics applications of the Integrated Tokamak Modelling (ITM-TF) tools to the
MHD stability analysis of experimental data and ITER scenarios},” in EPS
(2011), p. 2. \url{https://info.fusion.ciemat.es/OCS/EPS2011ABS/pdf/O2.103}

\bibitem{sol} L. S. Solov'ev,  JETP {\bf 26}, 400 (1968). \url{http://www.jetp.ras.ru/cgi-bin/dn/e_026_02_0400}

\bibitem{he} F. Herrnegger, ``{\it On the equilibrium and stability of the belt pinch}'', Proceedings of V. European Conference on Controlled Fusion and Plasma Physics, Grennoble, August 1972, Vol. I, p. 26 (1972).

\bibitem{ma} E. K. Maschke,  Plasma Phys. {\bf 15}, 535 (1973). \url{https://dx.doi.org/10.1088/0032-1028/15/6/006}

\bibitem{srila} R. Srinivasan, L. L. Lao and M. S. Chu,  Plasma Phys. Control. Fusion {\bf 52},  035007 (2010). \url{https://dx.doi.org/10.1088/0741-3335/52/3/035007}

\bibitem{maz} E. Mazzucato, Phys. Fluids {\bf 18}, 536 (1975). \url{
https://doi.org/10.1063/1.861186
}


\bibitem{gufr} L. Guazzotto and J. P. Freidberg, Phys. Plasmas {\bf 14}, 112508 (2007). \url{
https://doi.org/10.1063/1.2803759
}

\bibitem{cefr} Antoine J. Cerfon and Jeffrey P. Freidberg, Phys. Plasmas {\bf 17}, 032502 (2010). \url{
https://doi.org/10.1063/1.3328818
}

\bibitem{ceon} Antoine J. Cerfon and Michael O'Neil, Phys. Plasmas {\bf 21}, 064501 (2014). \url{
https://doi.org/10.1063/1.4881466
}

\bibitem{pfre} D. Pfirsch, E. Rebhan,  Nucl. Fusion {\bf 14}, 547 (1974). \url{https://dx.doi.org/10.1088/0029-5515/14/4/011}


\bibitem{cri} F. Crisanti, J. Plasma Phys. {\bf 85}, 905850210 (2019). \url{https://doi.org/10.1017/S0022377819000175}

\bibitem{ata} C. V. Atanasiu, S. G\"unter and K. Lackner, I.G.Miron, Phys. Plasmas,
{\bf 11}, 3510 (2004). \url{
https://doi.org/10.1063/1.1756167
}


\bibitem{mape} E. K. Maschke and H. Perrin,  Plasma Physics {\bf 22}, 579 (1980). \url{https://doi.org/10.1088/0032-1028/22/6/007} 

\bibitem{sol67} L.S. Solov'ev, in: Reviews of plasma physics, Vol. 3, ed.
M.A. Leontovich (Consultants Bureau, New York, 1967).

\bibitem{mosol} A.I. Morozov and L.S. Solov'ev, in: Reviews of plasma
physics, Vol. 8, ed. M.A. Leontovich (Consultants Bureau.
New York, 1980).

\bibitem{ham} E. Hameiri, Phys. Fluids {\bf 26}, 230 (1983). \url{https://doi.org/10.1063/1.864012}

\bibitem{ste} L. C. Steinhauer, Phys. Plasmas {\bf 6}, 2734 (1999). \url{https://doi.org/10.1063/1.873230}

\bibitem{clth} K. G. McClements and A. Thyagaraja, Plasma Phys. Control. Fusion {\bf 53},  045009 (2011). \url{https://doi.org/10.1088/0741-3335/53/4/045009}

\bibitem{il2001} V. I. Ilgisonis, Plasma Phys. Control. Fusion  {\bf 43},  1255 (2001). \url{https://doi.org/10.1088/0741-3335/43/9/307}

\bibitem{itra}  A. Ito, J. J. Ramos, N. Nakajima, Phys. Plasmas  {\bf 14}, 062502 (2007). \url{https://doi.org/10.1063/1.2741391}

\bibitem{kath2017} D. Kaltsas, G. N. Throumoulopoulos, P. J. Morrison,
Phys. Plasmas {\bf 24}, 092504 (2017). \url{https://doi.org/10.1063/1.4986013}

\bibitem{kath2018} D. Kaltsas, G. N. Throumoulopoulos, P. J. Morrison, 
J. Plasma Phys. {\bf 84}, 745840301 (2018). \url{https://doi.org/10.1017/S0022377818000338}

\bibitem{kath2019} D. Kaltsas, G. N. Throumoulopoulos, and P. J. Morrison, 
Physics of Plasmas {\bf 26}, 024501 (2019). \url{https://doi.org/10.1063/1.5080997}

\bibitem{cl}   R. A. Clemente, Nucl. Fusion {\bf 33},  963 (1993). \url{https://doi.org/10.1088/0029-5515/33/6/I12}

\bibitem{il1996} V. I. Ilgisonis, Phys. Plasmas {\bf 3}, 4577 (1996). \url{https://doi.org/10.1063/1.872074}

\bibitem{zwer} W. Zwingmann, L-G Eriksson and P.  Stubberfield, Plasma Phys. Control. Fusion {\bf 43},   1441 (2001). \url{https://doi.org/10.1088/0741-3335/43/11/302}

\bibitem{evth} A.  Evangelias, G. N. Throumoulopoulos,  Plasma Phys. Control. Fusion {\bf 58}, 045022 (2016). \url{https://doi.org/10.1088/0741-3335/58/4/045022}

\bibitem{kuev} Ap Kuiroukidis, A. Evangelias and G. N. Throumoulopoulos, Plasma Phys. Control. Fusion {\bf 59},  102001 (2017). \url{https://doi.org/10.1088/1361-6587/aa7c8d}

\bibitem{grad} H. Grad, Phys. Fluids {\bf 10}, 137 (1967). \url{https://doi.org/10.1063/1.1761965}

\bibitem{iabo} R. Iacono, A. Bondeson, F. Troyon, and R. Gruber, Phys. Fluids B {\bf 2}, 1794 (1990). \url{https://doi.org/10.1063/1.859451}

\bibitem{kamo2021} D. Kaltsas, G. N. Throumoulopoulos and P. J. Morrison, J. Plasma Phys. {\bf 87}, 835870502 (2021). \url{https://doi.org/10.1017/S0022377821000994}

\bibitem{kamo2023} D. Kaltsas, P. J. Morrison and G. N. Throumoulopoulos, J. Plasma Phys. {\bf 89}, 905890403 (2023). \url{https://doi.org/10.1017/S0022377823000557}

\bibitem{kaku2024} D. A. Kaltsas, A. Kuiroukidis, P. J. Morrison and G. N. Throumoulopoulos, Plasma Phys. Control. Fusion {\bf 66}, 065016 (2024). \url{https://doi.org/10.1088/1361-6587/ad4174}


\bibitem{tath}H. Tasso and G. N. Throumoulopoulos, Phys. Plasmas,
{\bf 5}, 2378 (1998). \url{
https://doi.org/10.1063/1.872912
}

\bibitem{sim} Ch. Simintzis, G. N. Throumoulopoulos, G. Pantis, H. Tasso,  Phys. Plasmas {\bf 8}, 2641 (2001). \url{
https://doi.org/10.1063/1.1371768
}

\bibitem{HELENA2} G. Poulipoulis, G. N. Throumoulopoulos, C. Konz, and ITM-TF Contributors, Phys. Plasmas {\bf 23}, 072507 (2016). \url{
https://doi.org/10.1063/1.4955326
}


\bibitem{kath2014} D. A. Kaltsas and G. N. Throumoulopoulos,
Phys. Plasmas {\bf 21}, 084502 (2014). \url{
https://doi.org/10.1063/1.4892380
}

\bibitem{shi2011} Bingren Shi,
Nucl. Fusion {\bf 51}, 023004 (2011). \url{https://dx.doi.org/10.1088/0029-5515/51/2/023004}

\bibitem{kuth2016} Ap. Kuiroukidis and G. N. Throumoulopoulos, Phys. Plasmas {\bf 23}, 114502 (2016). \url{https://doi.org/10.1063/1.4967346}

\bibitem{kaku2019} D. A. Kaltsas, A. Kuiroukidis and G. N. Throumoulopoulos,
Phys. Plasmas {\bf 26}, 124501 (2019). \url{https://doi.org/10.1063/1.5120341}

\bibitem{sal2018} A. R. Adem, S. M.  Moawad, Z. Naturforschung A {\bf 73}, 371 (2018) \url{https://doi.org/10.1515/zna-2017-0309 }

\bibitem{kuth2016a} Ap. Kuiroukidis and G. N. Throumoulopoulos, Phys. Plasmas {\bf 23}, 112508 (2016). \url{https://doi.org/10.1063/1.4968235}


\bibitem{kuth2014} Ap. Kuiroukidis and G. N. Throumoulopoulos, Phys. Plasmas {\bf 21}, 032509 (2014). \url{https://doi.org/10.1063/1.4869248}

\bibitem{poth2023} G. Poulipoulis, G. N. Throumoulopoulos, Phys. Plasmas  {\bf 30}, 114501 (2023). \url{https://doi.org/10.1063/5.0174091}

\bibitem{kath2016} D. A. Kaltsas and G. N. Throumoulopoulos,
Phys. Lett. A   {\bf 380}, 3373 (2016). \url{https://doi.org/10.1016/j.physleta.2016.08.011}

\bibitem{kuka2024} A. I. Kuiroukidis, D. A. Kaltsas, G. N. Throumoulopoulos,
    Phys. Plasmas {\bf 31}, 042503 (2024). \url{https://doi.org/10.1063/5.0198558}

\end{thebibliography}
\end{document}